\documentstyle[12pt]{article}
\begin{document}

\newcommand{\llq}{\lq\lq}
\newcommand{\rrq}{\rq\rq}
\newcommand{\rmd}{\mbox{d}}
\newcommand{\rme}{\mbox{e}}
\tolerance=5000

\def\pp{{\, \mid \hskip -1.5mm =}}
\def\cL{{\cal L}}
\def\be{\begin{equation}}
\def\ee{\end{equation}}
\def\bea{\begin{eqnarray}}
\def\eea{\end{eqnarray}}
\def\tr{{\rm tr}\, }
\def\nn{\nonumber \\}
\def\e{{\rm e}}
\def\D{{D \hskip -3mm /\,}}
\def\la{\label}
\def\PLB#1 {\Jl{Phys. Lett.}{#1B}}

\  \hfill
\begin{minipage}{3.5cm}
%September 2003 \\
\end{minipage}

%\vfill

\begin{center}
{\Large\bf Tachyon fields with effects of quantum matter in an
Anti-de Sitter Universe}

\vspace{6mm}

{\sc Emilio Elizalde}$^{\dagger}$\footnote{E-mail:
 elizalde@ieec.fcr.es} and

{\sc John Quiroga Hurtado}$^{\ddagger}$\footnote{E-mail:
jquiroga@utp.edu.co} \vspace{3mm}

{\sl $^\dagger$Consejo Superior de Investigaciones
Cient\'{\i}ficas
(ICE/CSIC) \\  Institut d'Estudis Espacials de Catalunya (IEEC) \\
Campus UAB, Fac Ciencies, Torre C5-Par-2a pl \\ E-08193 Bellaterra
(Barcelona) Spain} \vspace{3mm}

{\sl $^{\ddagger}$Department of Physics \\
Universidad Tecnol\'ogica de Pereira \\ Pereira, Colombia}

\vfill

{\bf Abstract}

\end{center}

We consider an Anti-de Sitter universe filled by quantum conformal
matter with the contribution from the usual tachyon and a perfect
fluid. The model represents the combination of a trace-anomaly
annihilated and a tachyon driven Anti-de Sitter universe. The
influence exerted by the quantum effects and by the tachyon on the
AdS space is studied. The radius corresponding to this universe is
calculated and  the effect of the tachyon potential is discussed, in
particular, concerning to the possibility to get an accelerated
scale factor for the  proposed model (implying an accelerated
expansion of the AdS type of universe). Fulfillment of the
cosmological energy conditions in the model is also
investigated.\vfill

\noindent PACS: 98.80.Hw,04.50.+h,11.10.Kk,11.10.Wx

\newpage

One of the most intriguing questions in today's Physics concerns the
nature of the dark energy present in the Universe we live in. The
existence of such energy, with almost uniform density distribution
and a substantial negative pressure, which completely dominates all
other forms of matter, is inferred from recent astronomical
observations \cite{SuNv}. In particular, according to recent
astrophysical data analysis, this dark energy seems to behave like a
cosmological constant, and it is responsible for the accelerating
expansion of the observable universe. And there are reasons to
believe that answering this question has much to do with the
possibility to explain the physics of the very early Universe.

Models of dark energy are abundant. One of the proposed candidates
for it is the phantom, thus called because it relies on a negative
energy field. The peculiar properties of a phantom scalar (with
negative kinetic energy) in a space with non-zero cosmological
constant have been discussed in an interesting paper by Gibbons
\cite{Gibbons}. It has been indicated there, that phantom properties
bear some similarity with quantum effects \cite{phtmtr}.
 The interesting property of the
investigation in \cite{Gibbons} is that it is easily generalizable
to other constant curvature spaces, as the Anti-de Sitter (AdS)
space. There is presently some interest in such spaces, coming in
particular from the AdS/CFT correspondence. According to it, the AdS
space might have in fact cosmological relevance \cite{cvetic}, e.g.
by way of increasing the number of particles created on a given
subspace \cite{jcap}. It could also be used to study a cosmological
AdS/CFT correspondence \cite{b494}: the study of a phantom field in
AdS space may give us a hint about the origin of such field via the
dual description. In the supergravity formulation, one may think of
the phantom as of a special RG flow for scalars in gauged AdS
supergravity. (Actually, such RG flow may correspond to an imaginary
scalar.)

Another candidate for dark energy is the tachyon. This is an
unstable field. The interest of models exhibiting a tachyon is
motivated by its role in the Dirac-Born-Infeld (DBI) action as a
description of the D-brane action \cite{copeland,s1,s2}. In spite of
the fact that the tachyon represents an unstable field, its role in
cosmology is still considered useful as a source of dark matter
\cite{Gibbons2} and, depending on the form of the associated
potential \cite{Paddy,Bagla,AF,AL,GZ}, it can lead to a period of
inflation. On the other hand, it is important to realize that a
tachyon with negative kinetic energy (yet another type of phantom)
can be introduced \cite{hao}. In that phantom/tachyon model $w$ is
naturally negative. In this case the late time de Sitter attractor
solution is admissible, and this is one of the main reasons why it
can be considered as an interesting model for the dark energy
\cite{hao}. Moreover, in order to understand the role of the tachyon
in cosmology it is necessary to study its effects on other
backgrounds, as in the case of an anti-de Sitter background.

In the present paper, we shall consider an AdS model filled with
classical matter, a perfect fluid, and a phantom/tachyon scalar,
taking also into account quantum contributions. The model can be
viewed as some generalized phantom/tachyon cosmology. In our theory,
quantum effects are described via the conformal anomaly, what is
reminiscent of the well known anomaly-driven inflation
\cite{starobinsky}. Such quantum effects are typical for the vacuum
energy (for a review, see \cite{book}). In special, we investigate
the analogies between our model formulated in AdS space and the
corresponding one formulated in a true de Sitter (dS) universe
\cite{snsdo}. By making use of the AdS/CFT correspondence, one can
expect that the tachyon (phantom/tachyon) field may emerge out of
some QFT instability in the dual description. It can originate as a
result of some phase transition. We will also study how the energy
conditions can indeed be fulfilled in a tachyonic AdS universe of
this sort, and what is the effect of the tachyon on the scale factor
and, correspondingly, on the accelerated inflation of such universe.

We start from an action for the tachyon, given by the following
expression
\be S_\phi=-\int{\rm d^4~\sqrt{-\rm g}
\left\{V(\phi)\sqrt{1+\lambda\rm
g^{\mu\nu}\partial_\mu\phi\partial_\nu\phi}+U(\phi)\right\}}\,.
\la{baction}\ee
This action describes an ordinary tachyon when $\lambda = 1$ and
$U(\phi)=0$, but if $\lambda = -1$ and $U(\phi)=0$ it corresponds to
a phantom/tachyon (see \cite{hao,snsdo}).

Let us now consider 4-dimensional Anti-de Sitter spacetime
(AdS$_4$), with the metric  chosen as \cite{brevik}
\be
 ds^2=e^{-2\lambda
\tilde{x_3}}(dt^2-(dx^1)^2-(dx^2)^2)-(d\tilde{x}^3)^2\,. \la{metr}
\ee
%\end{equation}
The simplest way to account for quantum effects (at least, for
conformal matter) is to include the contributions coming from the
conformal anomaly \be \label{OVII} T=b\left(F+{2 \over 3}\Box
R\right) + b' G + b''\Box R\ , \ee where $F$ is the square of 4d
Weyl tensor and $G$ the Gauss-Bonnet invariant, which are given by
\bea \label{GF} F&=&{1 \over 3}R^2 -2 R_{ij}R^{ij}+
R_{ijkl}R^{ijkl}\, , \nn G&=&R^2 -4 R_{ij}R^{ij}+ R_{ijkl}R^{ijkl}
\, . \eea
In general, with $N$ scalar, $N_{1/2}$ spinor, $N_1$ vector fields,
$N_2$ ($=0$ or $1$) gravitons and $N_{\rm HD}$ higher derivative
conformal scalars, $b$, $b'$ and $b''$ turn out to be \bea
\label{bs} && b={N +6N_{1/2}+12N_1 + 611 N_2 - 8N_{\rm HD} \over
120(4\pi)^2}\ ,\nn && b'=-{N+11N_{1/2}+62N_1 + 1411 N_2 -28 N_{\rm
HD} \over 360(4\pi)^2}\ , \nn &&  b''=0\ . \eea The contributions of
the conformal anomaly to $\rho$ and $p$ can be found in
\cite{NOev,NOOfrw}, namely \bea \label{hhrA3} \rho_A&=&-\left.{1
\over a^4}\right[b'\left( 6 a^4 H^4 + 12 a^2 H^2\right)
\\
&& + \left({2 \over 3}b + b''\right)\left\{ a^4 \left( -6 H
H_{,tt}- 18 H^2 H_{,t} + 3 H_{,t}^2 \right) + 6 a^2 H^2\right\}
\nn &&  -2b +6 b' -3b'' \Bigr] ,\nn \label{hhrAA1} p_A&=&b'\left\{
6 H^4 + 8H^2 H_{,t} + {1 \over a^2}\left( 4H^2 + 8 H_{,t}\right)
\right\} \nn && \left.+ \left({2 \over 3}b + b''\right)\right\{
-2H_{,ttt} -12 H H_{,tt} - 18 H^2 H_{,t} - 9 H_{,t}^2 \nn &&
\left. + {1 \over a^2} \left( 2H^2 + 4H_{,t}\right) \right\} - {
-2b +6 b' -3b''\over 3a^4} \ . \eea

The \llq radius\rrq of the Universe $a$ and the Hubble parameter
$H$ may be taken as
\be a\equiv L\rme^A\,,\quad H= \frac{1}{a}\frac{\rmd a}{\rmd t}=
\frac{\rmd A}{\rmd t}\,.\ee
Then, for such metric, the corresponding
 FRW equation has the following form
\bea H^2&=&-\frac{\kappa}{3}(\rho_\phi+\rho_A)\,,\la{frw}\\
\frac{\ddot{a}}{a}&=&
\frac{\kappa}{3}\left\{\frac{1}{2}\left(\rho_\phi+\rho_A\right)+\frac{3}{2}\left(p_\phi+p_A\right)\right\}\,.\la{frw1}\eea
Here $\rho_\phi$ and $p_\phi$ represent the tachyon energy density
and pressure, respectively.

From another side, by choosing $\phi$ to be depending only on
$x_3$, the tachyon action (\ref{baction}) takes the more simple
form
\bea
S_\phi=-\int\rmd^4xa^{-3}\left\{V(\phi)\sqrt{1-\lambda\dot{\phi}^2}+U(\phi)\right\}\,,\la{action}\eea
and, after varying (\ref{action}) with respect to $\phi$, the
resulting equation of motion for the tachyon is
\bea \lambda\ddot{\phi}+\left(\frac{V'(\phi)}{V(\phi)}-3\lambda
H\dot{\phi}\right)\left(1-\lambda\dot{\phi}^2\right)+\frac{U'(\phi)}{V(\phi)}\left(1-\lambda\dot{\phi}^2\right)^{3/2}=0\,.\la{eqmot}\eea

In the same way, after varying (\ref{action}) with respect to the
metric $g_{\mu\nu}$, we obtain for the energy density $\rho_\phi$
and pressure $p_\phi$:
\be
\rho_\phi=\frac{V(\phi)}{\sqrt{1-\lambda\dot{\phi}^2}}+U(\phi)\,\quad
p_\phi=-V(\phi)\sqrt{1-\lambda\dot{\phi}^2}-U(\phi)\,.\la{density}\ee
Assuming that the spacetime is Anti-deSitter for the scale factor
we have $a=\rme^{-\frac{x_3}{L}}$; thus, for the quantum energy density
and pressure on gets \cite{NOev, NOOfrw}
\be \rho_A=-p_A=-\frac{6b'}{L^4}\,.\la{q-energy}\ee
According to this, the FRW equations (\ref{frw}) and (\ref{frw1})
become
\bea
\frac{1}{L^2}&=&\frac{\kappa}{3}\left[\frac{V(\phi)}{2\sqrt{1-\lambda\dot{\phi}^2}}-\frac{3}{2}V(\phi)\sqrt{1-\lambda\dot{\phi}^2}-
U(\phi)+6b'\lambda^4\right]\,,\la{frw2}\\
\frac{1}{L^2}&=&-\frac{\kappa}{3}\left[\frac{V(\phi)}{\sqrt{1-\lambda\dot{\phi}^2}}+U(\phi)-6b'\lambda^4\right]\,.\la{frw3}\eea
Combining now (\ref{frw2}) and (\ref{frw3}), we obtain for $\dot{\phi}$
a trivial solution $\dot{\phi}^2=0$. Thus $\phi$ should be a
constant $\phi=\phi_o$, and therefore from eq. (\ref{frw3}) it
follows that
\be
L^{-2}=-\frac{\kappa}{3}\left(V(\phi_o)+U(\phi_o)-6b'\lambda^4\right)\,.\la{frw4}\ee
As we see, this solution differs from the one obtained in
\cite{snsdo} only on its sign, but this is no trivial change
since it will affect the
solutions for the tachyon potentials $V(\phi)$ and $U(\phi)$ in
our AdS spacetime.
Indeed, from eq. (\ref{eqmot}) it follows that
\be V'(\phi_o)+U'(\phi_o)=0\,.\la{eqmot1}\ee
Such a solution means that $V(\phi_o)+U(\phi_o)$ has an extremum
at $\phi=\phi_o$ or either that $V(\phi_o)+U(\phi_o)$ is a
constant. Let us now
look for a tachyon potential with $V(\phi)$ and $U(\phi)$
according to the proposal in \cite{kutasov} in the following form:
\bea
V(\phi)&=&V_o\left(1+\frac{\phi}{\phi_o\ln{\left(1+\frac{\phi}{\phi_o}\right)}}\right)\la{poten-v}\\
U(\phi)&=&-V_o\frac{\phi}{\phi_o\ln{\left(1+\frac{\phi}{\phi_o}\right)}}\,.\la{poten-u}\eea
It is easy to verify that $V(\phi)+U(\phi)=V_o=$ const, for any
value of $\phi$. On the other hand, as $\dot{\phi}=0$, we get
$\rho_{\phi}=-p_{\phi}=V(\phi_o)+U(\phi_o)$ and $w=-1$.
From (\ref{frw4}) the solution for $L^{-2}$ is found as
\be
L^{-2}=\frac{1}{4b'\kappa}\pm\sqrt{\frac{1}{16b'^2\kappa^2}+\frac{U_o}{6b'}}\,,\la{lambda2}\ee
where $U_o\equiv V(\phi_o)+U(\phi_o)$.
Since $b'$ is usually negative, we find real positive solutions
only if $U_o<0$. In fact we see that there is only one positive
solution taking the plus sign.
Thus, considering the case when $U_o$ is small enough, we find that
\be L^{-2}\sim\frac{-\kappa U_o}{3}\,.\ee
This solution corresponds to a universe which expands due only
to the tachyon perturbed by quantum effects. Therefore,
 in this scenario the inflationary regime emerging purely from
 quantum effects \cite{snsdo}
is not present in our AdS Universe as it occurs in the dS case
\cite{snsdo}.

On the other hand, taking into account that $\dot{\phi}=0$, we see
that our solutions satisfies
\be
w=\frac{p_{\phi}}{\rho_{\phi}}=\frac{p_{A}}{\rho_{A}}=\frac{p_{\phi}+p_{A}}{\rho_{\phi}+\rho_{A}}=-1\ee
Assuming $\dot{\phi}\neq 0$ in (\ref{density}), the effective
equation of state becomes
\be w=\frac{p_{\phi}}{\rho_{\phi}}=
-1+\frac{\lambda\dot{\phi}^2\left(V(\phi)
-\frac{\lambda\dot{\phi}^2U(\phi)}{\sqrt{1-\lambda\dot{\phi}^2}}\right)}{V(\phi)+U(\phi)\sqrt{1-\lambda\dot{\phi}^2}}\,.\ee
In this case, considering the situation of a phantom/tachyon
($\lambda<0$) for $V(\phi)>0$ and $U(\phi)\geq$, one finds that
$w\leq-1$.

Let us now look for solutions for the scale factor and the Hubble
parameter. We are interesting in their dependence
on the form choosen for the functions $V(\phi)$ and $U(\phi)$.
From (\ref{frw1}), (\ref{density}) and (\ref{q-energy}), the
solution for $H^2$ follows
\be
H^2=-\frac{\kappa}{3}\left[\frac{V(\phi)}{\sqrt{1-\lambda\dot{\phi}^2}}+U(\phi)-6b'L^{-4}\right]\,.\la{hubble}\ee
Then, for $\dot{\phi}=0$, we have
\be
H^2=-\frac{\kappa}{3}\left[V(\phi)+U(\phi)-6b'L^{-4}\right]=-\frac{\kappa}{3}\left(V_o-6b'L^{-4}\right)\,.\la{hubble1}\ee
This solution tells us that a non-imaginary scale factor is obtained
only under the condition $V_o<6b'L^{-4}$, and since $b'$ is negative,
 $V_o$ must be negative too.

Combining (\ref{frw}), (\ref{density}) and (\ref{q-energy}), the
following equation is obtained
\be
\frac{\ddot{a}}{a}=\frac{\kappa}{3}\left[\frac{V(\phi)}{2\sqrt{1-\lambda\dot{\phi}^2}}\left(3\lambda\dot{\phi}^2-2\right)-
U(\phi)+6b'L^{-4}\right]\,.\la{frw5}\ee
Now according to what was found before, we may consider the case when
$\dot{\phi}=0$; we get
\be
\frac{\ddot{a}}{a}=-\frac{\kappa}{3}\left[V(\phi)+U(\phi)+6b'L^{-4}\right]\,.\la{frw6}\ee
Equation (\ref{frw6}) may be rewritten as
\be \ddot{a}+\alpha_oa=0\,,\la{frw7}\ee
where $\alpha_o= -\frac{\kappa}{3}\left[V_o+6b'L^{-4}\right]$.

Taking into account the fact that $V_o$ must be negative, it is
possible to obtain a solution for $a$ depending on the sign of
$\alpha_o$, as follows
\bea a(x_3)=2a_o\cosh{(\alpha_ox_3)}\la{sol}\,.\eea

This solution (\ref{sol}) tells us that from our model the AdS
Universe is inflationary with an accelerating scale factor, since
$\ddot{a}>0$. Futhermore, we find that even when quantum effects
are small, the AdS Universe may still remain inflationary accelerated,
the inflation being induced only by the tachyon.

It is important to remark that the solutions obtained do not
depend explicitly on the form of the functions $V(\phi)$ and
$U(\phi)$ but only on the condition that
$V(\phi)+U(\phi)=V_o=$ const, for any value of $\phi$, according to
the form for the tachyon potential proposed in \cite{kutasov}.

An important issue to be investigated concerns the energy conditions
for this model, in other words, its consistency. In \cite{EEQJ}
the energy conditions were studied for an AdS universe with
phantoms and compared with those for the corresponding model in a
de Sitter universe. We want to know which of the energy
conditions can be fulfilled in the present model. The standard
ones in cosmology are the following:
\begin{enumerate}
\item Null Energy Condition (NEC): \be \label{phtm11} \rho + p
\geq 0. \ee \item Weak Energy Condition (WEC): \be \label{phtm8}
\rho\geq 0 \ \mbox{and}\ \rho + p \geq 0. \ee \item Strong Energy
Condition (SEC): \be \label{phtm9} \rho + 3 p \geq 0\ \mbox{and}\
\rho + p \geq 0. \ee \item Dominant Energy Condition (DEC): \be
\label{phtm10} \rho\geq 0 \ \mbox{and}\ \rho \pm p \geq 0. \ee
\end{enumerate}
In analogy with what was obtained when coupling the above model
with a usual phantom field
\cite{Gibbons} and the results for an AdS Universe
with a phantom and quantum matter \cite{EEQJ}, it is sensible to
 rewrite eqs.
(\ref{frw2}) and (\ref{frw3}) as
\bea
\frac{1}{L^2}&=&\frac{\kappa}{3}\left[\frac{V(\phi)}{2\sqrt{1-\lambda\dot{\phi}^2}}-\frac{3}{2}V(\phi)
\sqrt{1-\lambda\dot{\phi}^2}-U(\phi)+\frac{6b'}{L^4}\right.\nonumber\\ & &\left.-C^2+\frac{\rho_m}{2}+\frac{3}{2}p_m\right]\,,\la{frw9}\\
\frac{1}{L^2}&=&-\frac{\kappa}{3}\left[\frac{V(\phi)}{\sqrt{1-\lambda\dot{\phi}^2}}+U(\phi)-\frac{6b'}{L^4}-
\frac{C^2}{2}+\rho_m\right]\,.\la{frw10}\eea
Now, solving eqs. (\ref{frw9}) and (\ref{frw10}) for $\rho_m$ and
$p_m$, we obtain
\bea \rho_m&=& -\frac{3}{\kappa L^2}-
\frac{V(\phi)}{\sqrt{1-\lambda\dot{\phi}^2}}-U(\phi)+6b'L^{-4}+\frac{C^2}{2}\,,\la{density}\\
p_m&=& \frac{3}{\kappa L^2}+
V(\phi)\sqrt{1-\lambda\dot{\phi}^2}+U(\phi)-6b'L^{-4}+\frac{C^2}{2}\,.\la{pressure}\eea
% \vspace{3mm}
Let us consider with care the implications of these equations, i.e.
which are the restrictions on the anti-de Sitter cosmology which
follow from the energy conditions (\ref{phtm11}-\ref{phtm10}).

First, by combining (\ref{density}) and (\ref{pressure}), we
get
\bea \rho_m+p_m=
C^2-\frac{\lambda\dot{\phi}^2V(\phi)}{\sqrt{1-\lambda\dot{\phi}^2}}\,.\la{nec}\eea
From (\ref{nec}) we see that the NEC is satisfied only if
$V(\phi)>0\mbox{ and } \lambda<0$, i.e. for a phantom-like tachyon.
Now let us introduce
$\beta(\phi)\equiv\frac{C^2}{2}-\frac{V(\phi)}{\sqrt{1-\lambda\dot{\phi}^2}}-U(\phi)$.
Then, eq. (\ref{density}) is rewritten as follows
\bea \rho_m &=& \frac{\beta(\phi)}{L^4}
\left[L^2-\frac{3}{2\kappa\beta(\phi)}-\frac{1}{2}\sqrt{\left(\frac{3}{\kappa\beta(\phi)}\right)^2
-\frac{24b'}{\beta(\phi)}} \,\right]\nonumber\\ &\times&
\left[L^2-\frac{3}{2\kappa\beta(\phi)}+\frac{1}{2}\sqrt{\left(\frac{3}{\kappa\beta(\phi)}\right)^2
-\frac{24b'}{\beta(\phi)}}\, \right]\,.\la{density1}\eea
We  see that if $\beta(\phi)<0\,,\,\rho_m<0$, then the
WEC or DEC is not satisfied in such case. If $\beta(\phi)>0$,
since
$L^2-\frac{3}{2\kappa\beta(\phi)}+\frac{1}{2}\sqrt{\left(\frac{3}{\kappa\beta(\phi)}\right)^2
-\frac{24b'}{\beta(\phi)}}>0$, a non-trivial constraint on $L^2$
from WEC or DEC is obtained as follows:
\bea
L^2>\frac{3}{2\kappa\beta(\phi)}+\sqrt{\left(\frac{3}{2\kappa\beta(\phi)}\right)^2
-\frac{6b'}{\beta(\phi)}}\,.\la{l2}\eea
In addition, we have
\bea \rho_m+3p_m &=&
\frac{\gamma(\phi)}{L^4}\left[L^2-\frac{3}{\kappa\gamma(\phi)}-
\sqrt{\left(\frac{3}{\kappa\gamma(\phi)}\right)^2+\frac{12b'}{\gamma(\phi)}}\,\right]\nonumber\\
&\times& \left[L^2-\frac{3}{\kappa\gamma(\phi)}+
\sqrt{\left(\frac{3}{\kappa\gamma(\phi)}\right)^2+\frac{12b'}{\gamma(\phi)}}\,\right]\,.\eea
Here $\gamma(\phi)\equiv
2C^2+\frac{2-3\lambda\dot{\phi}^2}{\sqrt{1-\lambda\dot{\phi}^2}}V(\phi)+2U(\phi)$.
Then if $V(\phi)\,,\,U(\phi)>0$ and $\lambda<0$, we conclude that
$\gamma(\phi)>0$; on the other hand, if the quantity inside the square
root is negative,
\be
\left(\frac{3}{\kappa\gamma(\phi)}\right)^2+\frac{12b'}{\gamma}<0\,,\ee
we obtain $\rho_m + 3p_m>0$ and the SEC is satisfied.

On the other hand, if the quantity inside square root is positive,
from the SEC we obtain a non-trivial constraint on $L^2$, namely
\bea L^2 &<& \frac{3}{\kappa\gamma(\phi)}-
\sqrt{\left(\frac{3}{\kappa\gamma(\phi)}\right)^2+\frac{12b'}{\gamma(\phi)}}\nonumber\\
%&\vee& \nonumber\\L^2 &<& \frac{3}{\kappa\gamma(\phi)}+
%\sqrt{\left(\frac{3}{\kappa\gamma(\phi)}\right)^2
%+\frac{12b'}{\gamma(\phi)}}\,.
\eea
For the DEC we also have
\bea
\rho_m-p_m&=&-\frac{\eta(\phi)}{L^4}\left[L^2-\frac{3}{\kappa\eta(\phi)}-\sqrt{\left(\frac{3}{\kappa\eta(\phi)}\right)^2+
\frac{12b'}{\eta(\phi)}}\,\right]\nonumber\\ & \times &
\left[L^2-\frac{3}{\kappa\eta(\phi)}+\sqrt{\left(\frac{3}{\kappa\eta(\phi)}\right)^2+
\frac{12b'}{\eta(\phi)}}\,\right]\,,\eea
where
$\eta(\phi)=\frac{2-\lambda\dot{\phi}^2}{\sqrt{1-\lambda\dot{\phi}^2}}V(\phi)+2U(\phi)$.
In this case $V(\phi)\,,\,U(\phi)>0$, then $\eta(\phi)>0$ and from
the DEC we get the following constraint on $L^2$
\bea
\frac{3}{\kappa\eta(\phi)}-\sqrt{\left(\frac{3}{\kappa\eta(\phi)}\right)+\frac{12b'}{\eta(\phi)}}
< L^2 <
\frac{3}{\kappa\eta(\phi)}+\sqrt{\left(\frac{3}{\kappa\eta(\phi)}\right)+\frac{12b'}{\eta(\phi)}}\,.\eea

From the above analysis of the energy conditions, we are led to
the conclusion that they can always be fulfilled, provided
the constrains derived are imposed. In other words,
 these results lead to the possibility of the formation of an
 AdS universe out of quantum
matter effects and the presence of  dark energy, which in our
model is obtained from the contributions of the tachyon and
phantom fields. As a proof of consistency of the derivations
above, we may restrict the above analysis
for the energy conditions (and the constraints derived for $L^2$)
to the case when the contributions of the
tachyon are absent. Then we immediately see that the results
that we got in \cite{EEQJ} for an
anti-de Sitter Universe filled with quantum CFT with classical
phantom matter and a perfect fluid are recovered.

To summarize, a number of interesting conclusions can be drawn from
the study of the influence of tachyon/phantom and quantum effects in
an AdS universe. In particular, the possibility to give sense to
such a model, attending to the fact that the majority of the energy
conditions can be preserved ---within certain limits--- when matter
is composed of tachyon/phantom, perfect fluid and conformal matter.

In addition, it has been shown that our model represents a sort of
inflationary accelerated anti-de Sitter universe and that it still
may remain such, even if the quantum effects are not considered.

Furthermore, from our expressions we are able to recover
(\ref{sol}), i.e. the situation when the tachyon is not present, in
which case we again conclude that the anti-de Sitter Universe
remains accelerated ---knowing however that in this particular case
it will not be stable \cite{brevik,quiroga,quiroga1}.

\vspace{3mm}

\noindent {\bf Acknowledgments}

We are grateful to S.D. Odintsov for very helpful discussions. The
research of J.Q.H.  at UTP has been supported by a Professorship
from the Universidad Tecnol\'ogica de Pereira, Colombia. E.E.  has
been supported by DGICYT (Spain), project BFM2003-00620, by SEEU
grant PR2004-0126, and by CIRIT (Generalitat de Catalunya), grants
2002BEAI400019 and 2001SGR-00427.

\end{document}